\documentstyle[12pt]{article}

\begin{document}
\begin{titlepage}

\begin{center}
{ \large \bf

Effective abelian gauge theory for  lattice gluodynamics
and  chromoelectric string.  }
\end{center}

\vskip 3.0cm

\begin{center}

{ \large \bf  M.A.Zubkov $^{*}$} \\

\vskip 0.6cm

{\it Institute for Theoretical and Experimental Physics,  \\

B.Cheremushkinskaja, 25, Moscow, Russia}\\

\vskip 0.4cm

\end{center}

\vskip 4.0cm

\rm
\noindent

The lattice $SU(2)$  gluodynamics  in the maximal abelian projection
is reduced to the abelian theory, in which  the natural small
parameter exists.  We show that in the zeroth order of the expansion
in  this parameter the theory is equivalent to the compact abelian
gauge theory coupled to ghosts of charge 2.
 The theory is a
renormalizable and asymptotically free.  This theory is represented
  in the form of  the theory of open string  with the  boundary
consisting of the worldline of the quark. The ghosts live on
the worldsheet of the string. The naive continuum limit of such
string representation gives a simple expression for the
chromoelectric string action.  \vfill

\vskip 0.4cm

\begin{flushleft}
\noindent
\rule{5.1 in}{.007 in}\\
$^{*}${\small E-mail: $~~$ {\small \bf
 zubkov@vitep5.itep.ru  \\}}
\end{flushleft}

\end{titlepage}

\vfill\eject

\baselineskip 0.65cm

\sloppy

\section{Introduction}
The most important problem in the nonabelian theories is
the nonlinear selfinteraction of gluons. Due to this selfinteraction
the physical processes are not understood completely yet both
quantitatively and qualitatively.  t'Hooft suggested the method to
convert the nonabelean theory into the abelian one. We hope that
in this theory all  physical processes are  much more transparent.
This method is known as the procedure of abelian
projection\cite{THOP}.  There exists a lot of abelian projections, but
only in the one of them, in the so - called maximal abelian
projection, the monopoles are shown to be related to the dual
superconductor mechanism of confinement\cite{MAP}. This mechanism was
proposed by t'Hooft and Mandelstam \cite{Conf}.  The maximal abelian
projection makes the link variables as diagonal as possible. It
occurs that there exists a small parameter which is the ratio of the
contribution of the nondiagonal elements of link matrices  into the
effective action  to the contribution  of the diagonal
elements\cite{Eps}.  Lattice calculations show that this parameter is
less than  $0.1$ \cite{Eps}.

In this paper we show that in the zeroth order of the expansion in
this parameter the lattice $SU(2)$ theory becomes the compact abelian
gauge theory coupled to ghosts of charge 2. We can represent the
Wilson loop average in this theory as the sum over the
worldsheets of the electric string. The boundary of this string
consists of the given loop. The  ghost worldline is closed
and always lies on the string's worldsheet. The orientation of the
string worldsheet is changing on the  ghost's worldline.  The form
 of the string action gives the evidence of the existence of
confinement phase in this theory.  In the naive continuum limit of
the abelian projected theory we come to the asymptotically free
renormalizable theory. This shows that it's lattice version can admit
the continuum limit.

There is the  longstanding question in the chromodynamics -
what is the action of the chromoelectric string.  The investigation
of the simplest Nambu - Goto string showed, that it can not be
the physical string, because of the crumpling.
  A new form of the string action
   was recently
suggested, which possesses the
extended reparametrization invariance \cite{pol}. The string
representation of the Wilson loop average in this theory
satisfies the $SU(\infty)$ - loop equations without contact
terms.  In this paper we note that the string with this action
 arises as the naive continuum limit of the string representation of
 the effective abelian gauge theory mentioned above.

In   section  2 we consider the  gluodynamics
in  maximal abelian gauge.

In   section 3 we derive the effective abelian theory and the
string representation of the effective abelian theory.

In   section 4 we consider the naive continuum limit of the
effective abelian theory.

 \section{$SU(2)$ - gluodynamics in  the maximal abelian
gauge.}

We consider the $SU(2)$ - gluodynamics with  the Wilson action
$S(U) = \beta \sum_{plaq} (1-1/2 \mbox{Tr} U_{plaq})$. Here the sum is over
the plaquettes of the lattice. If the given plaquette consists of the
links $[xy]$,$[yz]$,$[zw]$,$[wx]$ then $U_{plaq} = U_{[xy]}
 U_{[yz]}  U_{[zw]}  U_{[wx]} $. We start
from the Wilson loop average

\begin{equation}
<W_C> = \int D U  \exp(- S(U)) \mbox{Tr} \Pi_{[xy]\in C} U_{[xy]}.
\end{equation}
Here the loop $C$ consists of links $[xy]$, $\Pi_{[xy]\in C}
U_{[xy]}$
is the ordered product of link matrices along the given loop.

The projection diagonalizes the link matrix
 $U$ as much as possible.
This procedure makes the functional $Q = \sum_{links} |U_{11}|^2$
maximal with respect to the gauge transformations.
 This gauge
condition is invariant under the Cartan subgroup of $SU(2)$. This
subgroup is created by the matrices
$diag(e^{i\alpha},e^{-i\alpha})$.
 We parametrize the matrix $U$ as $U_{11} =
cos(\rho) e^{i\theta}$, $U_{12} = sin(\rho) e^{i(\theta +
\lambda)}$. Then the field $\theta$ plays the role of abelian
gauge field. It transforms as $\theta_{xy} \rightarrow \theta_{xy}
+ \alpha_x - \alpha_y$ under the action of Cartan subgroup $U(1)$
of $SU(2)$. The infinitesimal elements of $SU(2)/U(1)$ can be
represented as
\begin{equation}
g = \left(\begin{array}{cc} 1 & \phi^+ \\
-\phi & 1\end{array} \right)\label{ph}
\end{equation}
where $\phi$ is complex - valued.

The differential condition for the maximal abelian gauge is
\begin{equation}
\frac{\delta}{\delta \phi_x} Q[U] = - \sum_{y,y>x}
U^{12}_{xy}\,U^{11}_{xy} + \sum_{y,y<x} U^{12}_{yx}(U^{11}_{yx})^*
= 0
\end{equation}

In order to fix the gauge we insert the unity into the functional
integral
\begin{equation}
 1 = \int Dg \delta(\frac{\delta}{\delta \phi_x} Q[U^g]) \Gamma (U^g)
  {\bf \rm Det} \, \frac{\delta^2}{\delta \phi_x \delta \phi_y}
  Q[U^g]
 \end{equation}
where $g \in SU(2)/U(1)$ and its infinitesimal element is given by
(\ref{ph}). The function $\Gamma$ is integer - valued and shares
out the field configurations from the set of solutions of the
differential condition mentioned above that are the global maxima
of the integral gauge condition.

There exists the small parameter in the theory, namely, the ratio
of the contribution into the effective action of nondiagonal part
of $U$ to the contribution  of the diagonal part of $U$.

We can rewrite the  Wilson loop average as follows.
\begin{eqnarray} <W_C> = \int DU D\bar\psi D\psi \exp(- \beta
\sum_{plaq}(1-1/2\mbox{Tr} U_{plaq}) \nonumber\\ +
\sum_{[xy]}\bar\psi_x [\frac{\delta^2}{\delta \phi_x \delta
\phi_y}
  Q[U^g]|_{g = 1}]\psi_y ) \Pi_{[xy]\in C} U_{[xy]}\nonumber\\
\delta(-\sum_{y,y>x} U^{12}_{xy} U^{11}_{xy} + \sum_{y,y<x}
U^{12}_{yx}(U^{11}_{yx})^* ) \Gamma(U).
\end{eqnarray}

Here the field $\theta$ is the abelian gauge field, $\psi$ are the
ghost anticommuting variables,
 and the
vector field of charge two
 $f = sin2\rho
\exp(i\lambda)$ is hidden in this expression.

\section{The effective abelian theory.}

As it was mentioned, there exists the small parameter in the maximal
 abelian projection.  Thus there exists the
   perturbation theory in this small parameter.
   In the zeroth
order of expansion in our small parameter,
 one can neglect the contribution of  the vector charged field.
 The direct calculation of $[\frac{\delta^2}{\delta \phi_x \delta
\phi_y}
  Q[U^g]|_{g = 1}]$
then leads us to the expression:

\begin{eqnarray}
<W_C> = \int D\theta D\bar\psi D\psi \exp(- \beta
\sum_{plaq}(1-cos(d\theta)_{plaq})
\nonumber\\ + \sum_{[xy]}(\bar\psi_x -
e^{2i\theta_{xy}}\bar\psi_y)(\psi_x - e^{-2i\theta_{xy}}\psi_y))+i(\theta,C)).
\end{eqnarray}
We write the Wilson loop average in the terms of some abelian
theory. We shall refer to it as to effective abelian theory. Here
and below we use the notations of differential forms on the
lattice (see \cite{BKT}).

In order to take into account the higher orders, one should
consider the full expression for  $\frac{\delta^2}{\delta \phi_x
\delta \phi_y}
  Q[U^g]$ and the function $\Gamma(U)$ which defines the fundamental modular region
\cite{fmr}.

Below we consider the leading approximation in  details.
  For simplicity we choose the Villain form of the action
of the effective abelian gauge theory.  It corresponds to the choice
of modified action for the initial $SU(2)$ gauge theory. One can
easily see that, for the case of the Wilson action, all the results
of this section remain valid, with some nonessential formal
complications.  For the Wilson loop we have

\begin{eqnarray}
<W_C> = \int D\theta D\bar\psi D\psi \sum_n \exp(- \beta
(d\theta + 2\pi n, d\theta + 2\pi n)
\nonumber\\ + \sum_{[xy]}(\bar\psi_x -
e^{2i\theta_{xy},}\bar\psi_y)(\psi_x - e^{-2i\theta_{xy}}\psi_y))+i(\theta,C)),
\end{eqnarray}

where $n$ is the integer - valued 1 - form.

Let us consider the duality transformation

\begin{eqnarray}
<W_C> = \int D\theta D\bar\psi D\psi \sum_n \exp(- \beta (d\theta
+ 2\pi n, d\theta + 2\pi n) \nonumber\\ + \sum_{[xy]}(\bar\psi_x -
e^{2i\theta_{xy},}\bar\psi_y)(\psi_x -
e^{-2i\theta_{xy}}\psi_y))+i(\theta,C)) \nonumber\\ = \int D\theta
DF \sum_{\delta j = 0} \sum_n   (-1)^{k(j)} A[j]
 \nonumber \\ \exp(- \beta (F,F) +i(\theta,C + 2j))\delta(F- d\theta -
2\pi n) \nonumber\\ = \int D\theta DF \sum_{\delta j =
0} \sum_\sigma
 (-1)^{k(j)}A[j] \nonumber\\  \exp(- \beta (F,F) + i(F-
d\theta , 2\pi \sigma) +i(\theta,C + 2j)) \nonumber\\ =
\sum_{\delta j = 0}(-1)^{k(j)} A[j] \sum_{\delta\sigma =
C+2j}\exp(-
 \frac{\pi^2}{\beta} (\sigma,\sigma)). \label{st}  \end{eqnarray}
Here $j$ is the integer - valued 1 - form representing the ghost
worldlines, $k(j)$ is the number of ghost loops, $A[j]$ is a
combinatorial factor, $\sigma$ is the integer -valued 2 -form
representing the electric flux strings.
 As it follows from the
condition $ \delta\sigma = C+2j $, the boundary of the string's
worldsheet consists of the given loop $C$ and the ghost worldline
$j$. The ghost worldline creates two strings. This fact we  can
treat as follows. The boundary of the string's worldsheet consists
of the given loop only. The ghosts live on the string's worldsheet
and the worldsheet change it's orientation on the ghost's worldline.

Here we represented the fermion determinant as the sum over closed
lines $j$.

 Thus,  we can see, that the
theory has the phase of confinement of the fundamental charges.

One can rewrite the expression  (\ref{st}) in the other way

  \begin{eqnarray} <W_C> = lim_{m\rightarrow \infty} \int D\theta
D\bar\psi D\psi \sum_n \exp(- \beta (d\theta + 2\pi n, d\theta + 2\pi
n) \nonumber\\ - \frac{4\pi^2\beta}{m^2}  (dn,dn)  +
\sum_{[xy]}(\bar\psi_x - e^{2i\theta_{xy},}\bar\psi_y)(\psi_x -
e^{-2i\theta_{xy}}\psi_y))+i(\theta,C)) \nonumber\\ =
lim_{m\rightarrow \infty}\sum_{\delta j
 = 0}(-1)^{k(j)} A[j] \int D B D \epsilon \sum_l \nonumber\\ \exp(-
 \frac{1}{4\beta} (dB + 2\pi *s[C+2j],dB + 2\pi
 *s[C+2j])\nonumber\\ - \frac{m^2}{4\beta} (d\epsilon + B + 2\pi
 l,d\epsilon + B + 2\pi l)).  \label{dsc}\end{eqnarray}
 This expression defines the dual representation of the effective
 abelian theory. Here   $B$ is the noncompact field, $\epsilon$ is
 the compact field, $\epsilon \in (-\pi,\pi]$, $s[C+2j]$ is the
surface, which boundary is $C+2j$, $l$ is an integer - valued one -
form.  We note, that the representation (\ref{st}) follows from the
expression (\ref{dsc}) when we apply the  BKT transformation
\cite{BKT}.

Let us note, that if we omit the summation over ghost worldlines
$j$, we come to the
noncompact Abelian Higgs theory with
infinitely deep potential \cite{BKT} and external monopole current
$C$.  In this theory $B$ is the gauge field, $\epsilon$ is the phase
of the monopole field: $\phi_{monop}= r \exp(i\epsilon)$.
Here the radial part $r$ of the monopole field  is frozen.
Due to the condition $m \rightarrow \infty$ we have: $r
\rightarrow \infty$.

Thus we treat the dual representation of the theory as
the dual superconductor theory with additional  monopole excitations,
which are the ghosts of the initial representation.  The
infiniteness of parameters of the potential is the deficiency of our
approximation.

\section{The naive continuum limit of the effective theory.}

The naive continuum limit of our theory is the QED with ghosts of
charge 2. The perturbative expansion of this theory differs from
the scalar electrodynamics by the sign $(-)$ for the ghost loops.
Due to this sign the model is assymptotic free. This gives us the
reason to believe, that the lattice effective abelian theory has
the  continuum
 limit (as we know,  the asymptotically free
nonabelian gauge theories have the continuum limit, while the compact
QED has not).

The naive continuum limit of the string representation of the theory in the
Villain form is the string theory with the  action:

\begin{eqnarray}
S = const \int d^2 \sigma_1 \int d^2 \sigma_2 \delta^{(4)}
(x(\sigma_1) - x(\sigma_2))  \\ \nonumber
\partial_{\alpha}x^{\mu}(\sigma_1)\partial_{\beta}x^{\nu}(\sigma_1)
\epsilon^{\alpha \beta}
\partial_{\gamma}x^{\mu}(\sigma_2)\partial_{\rho}x^{\nu}(\sigma_2)
\epsilon^{\gamma \rho}
\end{eqnarray}
Here the functions $x^{\mu}(\sigma)$ represent the space - time
coordinates of the string, $\sigma_1$ and $\sigma_2$ are the
coordinates, which parametrize the worldsheet of the string.  This
string possesses the extended reparametrization invariance \cite{pol}
and, thus, one can hope that it does not suffer from crumpling.
Also, it is known that the string theory with this action solves the
$SU(\infty)$ - loop equations without contact terms.  In this paper
we have found that the string theory with this action naturally
arises from the  effective abelian gauge theory.
The new feature is the existence of ghosts at the ends of the string.
The naive continuum limit of the expression for the Wilson loop
average (\ref{st}) has the form

\begin{eqnarray}
<W_C> = \int Dy  (-1)^{k([y(\tau)])} \int_{\delta [x(\sigma)] = C + 2
[y(\tau)]} D x \\ \nonumber \exp( - const_1 \int d^2 \sigma_1 \int d^2
\sigma_2 \delta^{(4)} (x(\sigma_1) - x(\sigma_2))\\ \nonumber
\partial_{\alpha}x^{\mu}(\sigma_1)\partial_{\beta}x^{\nu}(\sigma_1)
\epsilon^{\alpha \beta}
\partial_{\gamma}x^{\mu}(\sigma_2)\partial_{\rho}x^{\nu}(\sigma_2)
\epsilon^{\gamma \rho} \\  \nonumber
- const_2 \int d \tau \sqrt{(y')^2})
\end{eqnarray}
Here the functions $y^{\mu}(\tau) $ are the space - time coordinates
of  the worldline of ghosts, coordinate $\tau$ parametrizes this
worldline, $k([y])$ is the number of ghost loops.  The functional integrals
$\int D y $ and  $\int Dx$ are taken  over the coordinates of the
ghost worldline and over the coordinates
of the string worldsheet respectively. The condition $\delta
[x(\sigma)] = C + 2 [y(\tau)]$ can be treated as in the previous
section. The boundary of the worldsheet consists of the given loop
$C$.  The ghosts live on the worldsheet and change it's orientation.
  The measure $Dx$ is chosen in the
correspondence with the natural metric on the worldsheet of the
string $h_{\alpha \beta}(\sigma) =
\partial_{\alpha}x^{\nu}(\sigma)\partial_{\beta}x^{\nu}(\sigma)$.
The measure $Dy$  corresponds to the natural metric on the ghost
worldline $h(\tau) = \frac{d y^{\mu}}{d\tau}\frac{d y^{\mu}}{d\tau}$.
The ghosts just change the orientation of
the string's worldsheet. It influence the  string's action
only in the case, when one piece of the worldsheet is superimposed on
the other one. The entropy of this case seems to be negligible and
thus we treat the ghosts as unphysical objects. Hence, for example,
there are no reasons to consider the asymptotic states created by
them.

\section{Conclusions}
In this paper we derive the effective abelian theory for the $SU(2)$ -
gluodynamics and the corresponding  action for the chromoelectric
string.  This theory possesses the asymptotic freedom and, thus, we
hope that it's lattice  version has the continuum
limit.  The string action possesses the additional reparametrization
invariance and  we believe, that the crumpling problem is absent
for it.

The  theory obtained is shown to have the phase, in which the
fundamental charges are confined. Our approximation is the zeroth
order of the expansion in the small parameter existing in the
maximal abelian gauge of the theory.  We do not know what is
the behavior of this parameter when $\beta$ tends to $\infty$ where
we have the phase transition to the continuum theory. If this
parameter remains small and the
effective abelian theory has the continuum limit as $\beta$ tends to
$\infty$, than we can state that it is a selfconsistent approximation
to gluodynamics, which demonstrates the mechanism of confinement and
gives the reasonable expression for the chromoelectric string's
action.

One can easily see that our consideration remains valid with some
nonessential complications in the case of $SU(N)$ theory.

 The author is grateful to E.T.  Akhmedov, M.N.  Chernodub, F.V.
Gubarev, M.I.  Polikarpov, Yu.A.  Simonov  and  K.L.  Zarembo  for
useful discussions.  The work is partially supported by  grants
 INTAS-RFBR-95-0681, RFBR-96-02-17230a.

\end{document}